\documentclass[12pt]{article}
\usepackage{amsmath}
\usepackage{amssymb}
\usepackage{graphicx}
\usepackage{float}
\usepackage{indentfirst}
\usepackage{mathrsfs}
\usepackage{dsfont}

\newtheorem{theorem}{Theorem}[section]

\newtheorem{observation}{Observation}

\begin{document}
	\baselineskip 20pt
	\title{Robustness of $2 \times N \times M$ entangled states against qubit loss}
	\author{S.M. Zangi$^{a,b}$\;~and
		Cong-Feng Qiao$^{a}$\footnote{Corresponding author.}\\[0.5cm]
		{\small $a)$ Department of Physics, University of the Chinese Academy of Sciences}  \\
		{\small YuQuan Road 19A, 100049, Beijing, China}\\
		{\small $b)$ Department of physics, Southern University of Science and Technology}\\
		{\small 1088 Xueyuan Avenue, 518055 Shenzhen, China}\\
	}
	\date{}
	\maketitle
	
	\begin{abstract}
		Entanglement in quantum systems is usually degraded by interaction with the environment. From time to time, some parties of a multipartite entangled system may become decoherent with other parties of the system due to the interference with the environment. In this situation, it is interesting to know how much information the residual system would keep on carrying. In this paper, as a starting point for any entangled system, we investigate the property of the $2 \times N \times M$ state with qubit being disentangled, which is characterized by the measurement of robustness.
		
	\end{abstract}
	
	\section{Introduction}
	
	Entanglement \cite{EPR} is not only a key feature of quantum mechanics but also plays a main role in quantum supremacy \cite{arute19}.  People realize that this unique nature may be employed to accomplish some tasks unreachable within classical domain, such as quantum information and computation. So far there numerous unclear issues remain about the entanglement. To implement a quantum mission, normally a chain of entangled parties is required, rather the simplest two-body entangled state, of which the complex inner structures are still waiting for further exploration.
	Experiments involving quantum entanglement are generally impaired by environmental noises, for instance in quantum dots \cite{blv99},
	nuclear spins \cite{k98}, electronic spins \cite{v00}, optical lattice \cite{sm99}, and Josephson junctions \cite{usb01}. To characterize the multipartite entanglement inner property, Vidal and Tarrach \cite{vt99} introduced the measure of robustness to quantify the extent to which entangled states remain entangled under mixing. Subsequently, Schmidt robustness and the random Schmidt robustness were analogously introduced \cite{c06}, and the robustness was also defined in respect of partial loss \cite{dvc00,hs00}. Note that the concept of entanglement robustness against disentanglement of some parties makes sense only for the multipartite entanglement system. We call a tripartite entangled system robust while it remains entangled after the loss of one particle, otherwise it is referred to as fragile. It is found that the maximally entangled states violate the Bell inequality maximally \cite{Bell64}, meanwhile are maximally fragile \cite{gp08}. D\"ur {\it et al}. \cite{dvc00} considered the robustness of three-qubits GHZ and W states against the loss of a qubit, found that if any qubit of standard GHZ state is traced out the remaining biqubit reduced state will tend to be completely unentangled, which implies the GHZ is fragile. Whereas the other class of independent qutrit state, the W-state, behaves differently under the disposal of one qubit, which then says more robust than the GHZ state. Neven {\it et al}. \cite{nmb18} defined and discussed the robustness of multiqubit system with respect to the disentanglement of some qubits and classified the fragile and robust states under stochastic local operations and classical communication (SLOCC).
	
	When extending the two-level qubit to $N-$level qunit, one faces the problem that the $N$-dimensional generalization of the Pauli basis is hard to be Hermitian and unitary at the same time. In practice, qunit has many of applications in quantum communication and quantum computation, for instance in superconducting devices \cite{Reich14}.
	
	The main aim of this work is to extend the concept of robustness against particle loss to high dimensional systems. Note, throughout the paper the space dimensions $N$ and $M$ are taken to be finite by default.

	\section{Tripartite $2\times N \times M$ quantum state}
	
	By adopting the convention of Refs. \cite{clq10,lq13}, an arbitrary state of
	$2\times N \times M$ can be written as
	\begin{eqnarray}
	|\Psi_{2\times N \times M}\rangle =  \sum_{i,j,k} \gamma_{ijk}
	\, |i\rangle_{\psi_0} |j\rangle_{\psi_1} |k\rangle_{\psi_2}
	\label{psi1psi2}\; .
	\end{eqnarray}
	Here, the coefficient $\gamma_{ijk} \in \mathbb{C}$, characterizing the quantum state; $\psi_0$,
	$\psi_{1}$ and $\psi_{2}$ represent the qubit, qinits with dimension $N$ and $M$, respectively.
	Without loss of generality, hereafter we assume $N\leq M$. The entangled quantum state $2\times N \times M$ is then can be featured in a matrix form \cite{zlq17}, as
	\begin{eqnarray}
	|\Psi_{2\times N\times M}\rangle &=& \; \left(
	\begin{array}{cccc}
	\gamma_{111} & \gamma_{112} & \cdots & \gamma_{11M} \\
	\gamma_{121} & \gamma_{122} & \cdots & \gamma_{12M} \\
	\vdots & \vdots & \ddots & \vdots \\
	\gamma_{1N1} & \gamma_{1N2} & \cdots & \gamma_{1NM} \\  \hline
	\gamma_{211} & \gamma_{212} & \cdots & \gamma_{21M} \\
	\gamma_{221} & \gamma_{222} & \cdots & \gamma_{22M} \\
	\vdots & \vdots & \ddots & \vdots \\
	\gamma_{2N1} & \gamma_{2N2} & \cdots & \gamma_{2NM} \\
	\end{array}
	\right)
	=
	\begin{pmatrix}
	\Gamma_{\!1} \\
	\Gamma_{\!2} \\
	\end{pmatrix}\label{2mn-matrix} \; .
	\end{eqnarray}
	The matrix form of state $\Psi_{2\times N \times M}$ includes two submatrices,
	$\Gamma_{\{1,j,k\}}$ and $\Gamma_{\{2,j,k\}}$, which are both $N \times M$
	complex matrices.
	The density matrix $\rho = |\Psi_{2\times N \times M}\rangle\langle\Psi_{2\times N \times M}|$
	is then can be expressed in the submatrices form as
	\begin{equation}\label{dens_Operator}
	\rho = \left(
	\begin{array}{cc}
	\Gamma_1\otimes {\Gamma_1}^\dagger & \Gamma_1\otimes {\Gamma_2}^\dagger \\
	\Gamma_2\otimes {\Gamma_1}^\dagger & \Gamma_2\otimes {\Gamma_2}^\dagger \\
	\end{array}
	\right)\;,
	\end{equation}
	where every submatrix $\Gamma\otimes{\Gamma}^\dagger$ has dimension
	$NM\times NM$, and the dimension of $\rho$ is $2NM\times 2NM$.
	
	\subsection{The properties of $2\times N \times M$ state with the loss of qubit}
	
	Of the tripartite state $|\psi\rangle_{2\times m\times n}$, if the qubit is about to decouple from other two particles, the residual  state $\rho_{AB}$ can be expressed as $\rho_{\neg{\psi_0}} = \mathrm{Tr}_{\psi_{0}}[|\Psi\rangle \langle \Psi|]$.
	The entangled state $|\Psi\rangle_{2\times m\times n}$ is said
	fragile (vs robust) with the loss of qubit if $\rho_{\neg{\psi_0}}$ is separable (vs entangled) \cite{nmb18}.
	The residual bipartite dimensional $N\times M$ state in Bloch representation can be expressed as \cite{lq18a,lq18b}:
	\begin{eqnarray}
	\rho_{AB} & = & \frac{1}{NM} \mathds{1} \otimes \mathds{1} + \frac{1}{2M} \vec{a} \cdot \vec{\lambda}\otimes \mathds{1} + \frac{1}{2N} \mathds{1} \otimes \vec{b} \cdot \vec{\sigma} + \frac{1}{4} \sum_{\mu=1}^{N^2-1} \sum_{\nu=1}^{M^2-1} \mathcal{T}_{\mu\nu} \, \lambda_{\mu} \otimes \sigma_{\nu} \;. \label{rhoAB-Bloch-Gen}
	\end{eqnarray}
	Here, $\mathrm{Tr}[{\rho}_{AB}^2]<1$, which means (\ref{rhoAB-Bloch-Gen}) can only be in mixed state; $\mathds{1}$ denotes identity matrix; the components of $\vec{a}$ and $\vec{b}$ are respectively
	$a_{\mu} = \mathrm{Tr}[\rho_{AB} (\lambda_{\mu}\otimes \mathds{1})]$ and $b_{\nu} = \mathrm{Tr}[ \rho_{AB} (\mathds{1} \otimes \sigma_{\nu})]$;
	and the correlation matrix $\mathcal{T}_{\mu\nu} = \mathrm{Tr}[\rho_{AB}(\lambda_{\mu} \otimes \sigma_{\nu})]$.
	The vector $\vec{\lambda}$ in equation (\ref{rhoAB-Bloch-Gen}) is defined as
	$\vec{\lambda} \equiv (\lambda_1, \cdots, \lambda_{N^2-1})^{\mathrm{T}}$ with $\lambda_{\mu}$
	being the SU($N$) group generators, and $\vec{\sigma}$ is defined analogously. Explicitly, for example, in the case of qubit the three
	SU(2) generators are Pauli matrices, and for $N=3$ , the case of qutrit, generators are eight Gell-Mann matrices.
	
	A mixed bipartite state with quantum correlation takes the following form:
	\begin{equation}
	\rho_{AB} = \sum_{i=1}^L p_i \rho_i^{(A)} \otimes \rho_i^{(B)} \;, \label{Gen-Sep-exp}
	\end{equation}
	where $\rho_i^{(A)}$ and $\rho_i^{(B)}$ are the density matrices of the two parties, and the Bloch forms of individual subsystems are  $\rho^{(A)}_i = \displaystyle \frac{1}{N}\mathds{1} + \frac{1}{2}  \vec{r}_i \cdot \vec{\lambda}$ and $\rho^{(B)}_j = \displaystyle \frac{1}{M} \mathds{1} + \frac{1}{2} \vec{s}_j \cdot \vec{\sigma}$. The vectors $\mathbf{r}=\left(r_{1}, r_{2} \cdots r_{N^{2}-1}\right)^{T} \in \mathbb{R}^{N^{2}-1}$ and $\mathbf{s}=\left(s_{1}, s_{2} \cdots s_{M^{2}-1}\right)^{T} \in \mathbb{R}^{M^{2}-1}$
	which completely characterize the density operators are called Bloch vectors or coherence vectors. The norms of these vectors describe the mixedness (or quantumness) of states \cite{lq18a}, i.e.,
	\begin{equation}
	\|\mathbf{r}\|_{2}\leq\sqrt{\frac{2(N-1)}{N}}\;~~~,~~\|\mathbf{s}\|_{2}\leq\sqrt{\frac{2(M-1)}{M}}\;,   \label{norm-vec}
	\end{equation}
	where $\|\cdot\|_{2}$ represents the Euclidean norm on $\mathbb{R}^{N^{2}-1}$ and equality hold only for pure states.
	By comparing Eq. (\ref{Gen-Sep-exp}) with Eq. (\ref{rhoAB-Bloch-Gen}), we get
	\begin{eqnarray}
	\sum_i p_i\vec{r}_i = \vec{a} \;  , \; \sum_j p_j \vec{s}_j = \vec{b} \; , \;
	\sum_{k=1}^{n} p_k \vec{r}_k \vec{s}_k^{\,\mathrm{T}} = \mathcal{T} \; . \label{General-form-density}
	\end{eqnarray}
	Here, the subscripts in $\vec{r}_i$, $\vec{s}_j$ label different Bloch vectors
	and the correlation matrix $\mathcal{T}$ has the matrix elements $\mathcal{T}_{\mu\nu}$.
	The Eq. (\ref{General-form-density}) can be expressed in matrix form, as
	\begin{equation}
	M_{r} \vec{p}=\vec{a}\;,~~~ M_{s} \vec{p}=\vec{b}\;,~~~ M_{r p} M_{s p}^{\mathrm{T}}=\mathcal{T}\;.\label{sep-matrix}
	\end{equation}
	In (\ref{sep-matrix}) the matrix $M_{r}$ composed of $\vec{r}_i$ is in  dimension $(N^2-1)\times L$, and $M_{s}$ composed of $\vec{s}_j$ is in dimension $(M^2-1)\times L$; $\vec{p}=\left(p_{1}, \ldots, p_{L}\right)^{\mathrm{T}}$ and $M_{r p}=M_{r} D_{p}^{\frac{1}{2}}$, $M_{s p}=M_{s} D_{p}^{\frac{1}{2}}$ with
	$D_{p}=\operatorname{diag}\left\{p_{1}, p_{2}, \ldots, p_{L}\right\}$.

	The local density matrices for particles $A$ and $B$ are then obtained by using partial trace as
	\begin{equation}
	\rho_A = \mathrm{Tr}_B[\rho_{AB}] = \frac{1}{N} \mathds{1} + \frac{1}{2}\vec{a} \cdot \vec{\lambda} \; , \; \rho_B = \mathrm{Tr}_A[\rho_{AB}] = \frac{1}{M} \mathds{1} + \frac{1}{2}\vec{b} \cdot \vec{\sigma} \; ,
	\end{equation}
	where $\vec{a}$ and $\vec{b}$ are Bloch vectors of local density matrices of the one-particle quantum states.
	Since reduced density matrices are Hermitian, and hence can be unitarily diagonalized as
	\begin{align}
	\rho'_A & = U_A \rho_{A} U_A^{\dag} = \mathrm{diag}\{\lambda_1^{(A)},\cdots, \lambda_n^{(A)}, 0, \cdots,0\}\; , \label{diag-rhoa} \\ \;
	\rho'_B & = U_B \rho_{B} U_B^{\dag} = \mathrm{diag}\{\lambda_1^{(B)},\cdots,\lambda_m^{(B)}, 0, \cdots,0\} \; , \label{diag-rhob}
	\end{align}
	where $\lambda_i^{(A)}$ and $\lambda_i^{(B)}$ are positive real numbers and $n,m$ are local ranks of $\rho_A$ and $\rho_B$ respectivelly that may not be fully ranked for $n<N$ and $m<M$.
	As entanglement is invariant under unitary operators, so the states $\rho_{A B}^{\prime}=(U_{A} \otimes U_{B})\rho_{A B}(U_{A}^{\dagger} \otimes U_{B}^{\dagger})$ and
	$\rho_{AB}$ have the same strength of entanglement.
	
	\begin{theorem}
		If the qubit drops off from a robust quantum state $|\psi_{2\times N\times M}\rangle$, the residual bipartite state
		$\rho_{AB}$ with local ranks $n<N$ and $m<M$ must be an irreducible $n\times m$ bipartite state with full rank.
		\label{theorem-1}
	\end{theorem}
	
	\noindent{\bf Proof:} If a quantum state $|\psi_{2\times N\times M}\rangle$ is robust against the qubit loss, its bipartite
	density matrix $\rho_{AB}$ should remain to be entangled. Suppose $\rho'_{AB}$ is separable then
	$\rho'_{AB} = \sum_{i} p_i\rho_{i}^{(A)} \otimes \rho_{i}^{(B)}$, and $\rho'_A$ and $\rho'_B$ in (\ref{diag-rhoa}) and (\ref{diag-rhob}) read respectively:
	\begin{equation}
	\rho'_A = \sum_{i}p_i \rho_i^{(A)} \; , \;  \rho'_B = \sum_{i}p_i \rho_i^{(B)} \;.
	\end{equation}
	Here $p_i>0$ with $\sum_i p_i=1$; $\rho'_A$, $\rho_i^{(A)}$, $\rho'_B$, and $\rho_{i}^{(B)}$ are all positive semidefinite matrices.
	According to equation (\ref{diag-rhoa}), we notice that $\rho_i^{(A)}$ obviously can only take the following form
	\begin{equation}
	\rho_i^{(A)} =
	\begin{pmatrix}
	X_{n\times n} & 0 \\
	0 & 0
	\end{pmatrix}_{N\times N}\;  ,
	\end{equation}
	which in form of the Bloch vectors $\vec{r}_i$ of $\rho_i^{(A)}$ writes
	\begin{equation}
	\rho_i^{(A)} = \left(\frac{1}{n} \mathds{1} + \frac{1}{2}\sum_{\mu =1}^{n^2-1} r_{i\mu} \lambda_{\mu} \right)_{n\times n} \oplus \mathbf{0}_{(N-n) \times (N-n)} \;\; .
	\end{equation}
	Here $r_{i\mu}$ are components of $\vec{r}_i$ which lie in the Bloch vector space of SU(n) $\subset$ SU(N).
	Similar argument applies to $\rho_i^{(B)}$ as well. That means $\rho'_{AB}$ is a separable mixed state
	with $\mu\leq n^2-1$ and $\nu\leq m^2-1$, i.e., it is a reducible $n\times m$ state. Q.E.D.
	
	According to theorem (\ref{theorem-1}), we need only to consider the entanglement robustness for those mixed states whose reduced
	density matrices have full local ranks. Whereas, according to Ref.\cite{normal-form} the full local rank state
	could be further transformed into the normal form with maximally mixed subsystems.
	The normal form is entangled if and only if the original state is entangled. As a maximally mixed state has null Bloch vectors, the normal form of a bipartite state $\rho_{AB}$ may be obtained
	by taking $\vec{a}=0$ and $\vec{b}=0$ in equation (\ref{rhoAB-Bloch-Gen}), like
	\begin{equation}
	\rho_{A B} \mapsto \widetilde{\rho}_{A B}=\frac{1}{N M}  \mathds{1} \otimes  \mathds{1}+\frac{1}{4} \sum_{\mu=1}^{N^{2}-1} \sum_{\nu=1}^{M^{2}-1} \widetilde{\mathcal{T}}_{\mu \nu} \lambda_{\mu} \otimes \sigma_{\nu}\;.
	\end{equation}
	Due to above discussion, hereafter, reduced density operators $\rho_{AB}$ are assumed to be in their normal forms. It is noteworthy that the separability of the system in normal form was discussed as well in  Refs. \cite{separa-normal-form, normal-forms-criteria}.

	Since equations in (\ref{sep-matrix}) constrain the correlation matrix of separable state and its components, by applying the singular value decomposition (SVD) on the correlation matrix we have
	\begin{equation}
	\mathcal{T}=\left(\vec{u}_{1}, \ldots, \vec{u}_{N^{2}-1}\right) \Lambda_{\tau}\left(\vec{v}_{1}, \ldots, \vec{v}_{M^{2}-1}\right)^{\mathrm{T}} \; , \label{svdcorrelation}
	\end{equation}
	which can be expressed in a compact form $\mathcal{T}=\sum_{\mu=1}^{l} \tau_{\mu} \vec{u}_{\mu} \vec{v}_{\mu}^{\mathrm{T}}$ since $\Lambda_{\tau}$ has rank $l$. For $l$ nonzero values of $\tau_{\mu}$,
	the corresponding left and right singular vectors span a dimension $l$ subspace in Bloch space. $\mathcal{S}_{l}^{(A)} \equiv\operatorname{span}\left\{\vec{u}_{1}, \ldots, \vec{u}_{l}\right\} \subseteq \mathcal{S}_{N^{2}-1}$ and $\mathcal{S}_{l}^{(B)} \equiv \operatorname{span}\left\{\vec{v}_{1}, \ldots, \vec{v}_{l}\right\} \subseteq \mathcal{S}_{M^{2}-1}$.
	The singular value matrix $D_{\tau}=\operatorname{diag}\left\{\tau_{1}, \ldots, \tau_{l}, 0, \ldots, 0\right\}$ is a $L \times L$ diagonal matrix.
	
	Applying SVD on $M_{r p}$ and $M_{s p}$ we have
	\begin{equation}
	M_{r p}=R^{(1)} \Lambda_{\alpha} Q^{(1)},~~~ M_{s p}=R^{(2)} \Lambda_{\beta} Q^{(2)} \;. \label{svdfactor}
	\end{equation}
	Here $R^{(1)} \in \mathrm{SO}\left(N^{2}-1\right)$, $R^{(2)} \in \mathrm{SO}\left(M^{2}-1\right)$, and $Q^{(1)}, Q^{(2)} \in \mathrm{SO}(L)$. The singular value matrix of $M_{r p}$ writes $\Lambda_{\alpha}=\left({D_{\alpha}}, {0}\right)^{T} \in \mathbb{R}^{\left(N^{2}-1\right) \times L}$ if $N^{2}-1>L$, while
	$D_{\alpha}=\left(\Lambda_{\alpha}, {0}\right)^{T} \in \mathbb{R}^{L \times L}$ if $N^{2}-1<L$, with
	$D_{\alpha}=\operatorname{diag}\left\{\alpha_{1}, \ldots, \alpha_{L}\right\}$. Similarly we can get
	$\Lambda_{\beta}$ and $D_{\beta}$. Let $\mathcal{T}$ has $l$ nonzero singular values satisfying $\tau_{1} \geq\tau_{2} \geq \cdots \geq \tau_{l}>0$, $M_{r p}$ then has $n$ nonzero singular values $\alpha_{1} \geq \alpha_{2} \geq \cdots \geq \alpha_{n}>0$ and
	$M_{s p}$ has $m$ nonzero singular values $\beta_{1} \geq \beta_{2} \geq \cdots \geq \beta_{m}>0$.
	Then from the decomposition $\mathcal{T}=M_{r p} M_{s p}^{\mathrm{T}}$ we obtain the Sylvester's rank inequality:
	$(n+m-L) \leq l \leq \min \{n, m\} \leq \max \{n, m\} \leq L$.
	
	According to Theorem 1 of Ref.\cite{lq18a},
	$M_{r p}$ and $M_{s p}$ in Eq. (\ref{sep-matrix}) can be also expressed as
	\begin{eqnarray}
	M_{r p} &=& M_{r} D_{p}^{\frac{1}{2}}=\left(\vec{u}_{1}, \ldots, \vec{u}_{L}\right) X D_{\alpha} Q^{(1)}\;,\label{svd-mrp}\\
	M_{s p} &=& M_{s} D_{p}^{\frac{1}{2}}=\left(\vec{v}_{1}, \ldots, \vec{v}_{L}\right) Y D_{\beta} Q^{(2)}\;,\label{svd-msp}
	\end{eqnarray}
	where $X, Y, Q^{(1,2)}$ are orthogonal matrices, $\vec{u}_{\mu}$ and $\vec{v}_{\nu}$ are the left and right singular
	vectors of $\mathcal{T}$, respectively. $D_{\alpha} Q^{(1)} Q^{(2) \mathrm{T}} D_{\beta}^{\mathrm{T}}$ has same singular values as $D_{\tau}$.
	
	\begin{theorem}
		When column vectors $\overrightarrow{r_{i}}$ and $\overrightarrow{s_{j}}$ surpass the Bloch vectors of density matrix in length, the reduced quantum state $\rho_{AB}$ is then entangled.
		\label{theorem-2}
	\end{theorem}
	\noindent{\bf Proof:} According to SVD, the correlation matrix of a separable state reads
	\begin{equation}
	\mathcal{T}=\left(\vec{u}_{1}, \ldots, \vec{u}_{L}\right) D_{\tau}\left(\begin{array}{c}{\vec{\nu}_{1}^{\mathrm{T}}} \\ {\vdots} \\ {\vec{\nu}_{L}}\end{array}\right)\; .
	\end{equation}
	From the conclusion of Eqs. (\ref{svd-mrp}) and (\ref{svd-msp}),
	\begin{equation}
	\left(\begin{array}{ccccc}
	{\tau_{1}} & {0} & {\cdots} & {0} & {0} \\
	{0} & {\tau_{2}} & {\cdots} & {0} & {0} \\
	{\vdots} & {\vdots} & {\ddots} & {\vdots} & {\vdots} \\
	{0} & {0} & {\cdots} & {\tau_{l}} & {0} \\
	{0} & {0} & {\cdots} & {0} & {0}\end{array}\right)=
	\left(\begin{array}{ccccc}
	{\alpha_{1}} & {0} & {\cdots} & {0} & {0} \\
	{0} & {\alpha_{2}} & {\cdots} & {0} & {0} \\
	{\vdots} & {\vdots} & {\ddots} & {\vdots} & {\vdots} \\
	{0} & {0} & {\cdots} & {\alpha_{l}} & {0} \\
	{0} & {0} & {\cdots} & {0} & {0}\end{array}\right)Q Q^T
	\left(\begin{array}{ccccc}
	{\beta_{1}} & {0} & {\cdots} & {0} & {0} \\
	{0} & {\beta_{2}} & {\cdots} & {0} & {0} \\
	{\vdots} & {\vdots} & {\ddots} & {\vdots} & {\vdots} \\
	{0} & {0} & {\cdots} & {\beta_{l}} & {0} \\
	{0} & {0} & {\cdots} & {0} & {0}\end{array}\right)\; ,\label{svdT}
	\end{equation}
	where, $Q \in \mathrm{SO}(l+1)$ with elements $Q_{(l+1) j}=\sqrt{p}_{j} ; p_{j} \geq 0$ in the last row and $\sum_{j=1}^{l+1} p_{j}=1$.
	Let $\alpha_{i}=\left(\frac{2}{N(N-1)}\right)^{\frac{1}{2}} \sqrt{\kappa_{i}}, \beta_{i}=\left(\frac{2}{M(M-1)}\right)^{\frac{1}{2}} \sqrt{\kappa_{i}},$
	and $\kappa_{i}=\tau_{i} \frac{\sqrt{N(N-1) M(M-1)}}{2},$ we then have
	\begin{equation}
	\mathcal{K} \equiv \sum_{i=1}^{l} \kappa_{i}=\frac{\sqrt{N(N-1) M(M-1)}}{2} \sum_{j=1}^{l} \tau_{i} \leq 1 \; .\label{sumk}
	\end{equation}

	According to Ref.\cite{vi07}, for separable state, $\tau_{i}=\alpha_{i} \beta_{i}$. Comparing Eq.(\ref{svdT})
	with $\mathcal{T}=M_{r p} M_{s p}^{\mathrm{T}}$, the Bloch vectors $\overrightarrow{r_{j}}$ and $\overrightarrow{s_{j}}$ may be obtained, i.e.,
	\begin{equation}
	\sqrt{p_{j}} \overrightarrow{r_{j}}=\left(\alpha_{1} Q_{1 j}, \alpha_{2} Q_{2 j}, \ldots, \alpha_{l} Q_{l j}\right)^{\mathrm{T}}\; ,
	\end{equation}
	\begin{equation}
	\sqrt{p}_{j} \overrightarrow{s_{j}}=\left(\beta_{1} Q_{1 j}, \beta_{2} \mathrm{Q}_{2 j}, \ldots, \beta_{l} \mathrm{Q}_{l j}\right)^{\mathrm{T}}\; .
	\end{equation}
	Therefore, the norms of column vectors $\overrightarrow{r_{i}}$ and $\overrightarrow{s_{i}}$ are
	\begin{equation}
	p_{j}\left|\vec{r}_{j}\right|^{2}=\sum_{i=1}^{l} \alpha_{i}^{2} Q_{i j}^{2}=\frac{2}{N(N-1)} \sum_{i=1}^{l} \kappa_{i} Q_{i j}^{2}\ ,\;\label{rjnorm}
	\end{equation}
	\begin{equation}
	p_{j}\left|\vec{s}_{j}\right|^{2}=\sum_{i=1}^{l} \beta_{i} Q_{i j}^{2}=\frac{2}{M(M-1)} \sum_{i=1}^{l} \kappa_{i} Q_{i j}^{2}\;\label{sjnorm} .
	\end{equation}
	Set the probability distribution $p_{j}=\frac{1}{\mathcal{K}} \sum_{i=1}^{l} \kappa_{i} Q_{i j}^{2}$, then from Eqs. (\ref{rjnorm} ) and (\ref{sjnorm}) we get for separable states
	\begin{equation}
	\left|\overrightarrow{r_{j}}\right|^{2}=\frac{2 \mathcal{K}}{N(N-1)} \leq \frac{2}{N(N-1)},~~~\left|\vec{s}_{j}\right|^{2}=\frac{2 \mathcal{K}}{M(M-1)} \leq \frac{2}{M(M-1)}\; .
	\end{equation}
	This indicates that, of the entangled states, column vectors $\overrightarrow{r_{i}}$ and $\overrightarrow{s_{j}}$ surpass the Bloch vectors of density matrix in lengths. Q.E.D.
	
	The matrix unfolding in tensor also names matrization of tensor \cite{lmv00}. We can define the Ky Fan norm of the
	$N-$order tensor $\mathcal{T}^{N}$ over $N$ matrix unfolding as
	\begin{equation}
	\left\|\mathcal{T}^{(N)}\right\|_{K F}=\max \left\{\left\|T_{(n)}^{(N)}\right\|_{K F}\right\}, n=1, \ldots, N\;.\label{kyfan-norm}
	\end{equation}
	Here, $\left\|\mathcal{T}^{(N)}\right\|_{K F}$ is the Ky Fan norm of matrix $T_{(n)}^{(N)}$
	defined as the sum of singular values of $T_{(n)}^{(N)}$ \cite{Matrix-analysis}.
	For the normal form of bipartite state with maximally mixed subsystems, if
	\begin{equation}
	\|\mathcal{T}\|^2_{K F} >\frac{4(N-1)(M-1)}{N M}\; ,\label{separable-kfnorm}
	\end{equation}
	the state is entangled \cite{lq18b}.
	
	The robustness (resp. fragility) of a multipartite entangled state is a LU-invariant,
	so the state $\rho'_{AB} = (U_A\otimes U_B) \rho_{AB} (U_A^{\dag} \otimes U_B^{\dag})$
	has the same robustness as $\rho_{AB}$. Now we are capable of recognizing
	the robust and fragile tripartite state against qubit loss with reducible bipartite states.
	
	{\bf Example 1:} Application to $2\times 2\times 4$ state.
	
	If the qubit in $2\times 2\times 4$ is decouple from the system due to some reason, the residual state $2\times 4$ would be a mixed, written as:
	\begin{equation}
	\rho_{A B}=\frac{1}{2 \cdot 4} \mathds{1}\otimes \mathds{1}+\frac{1}{4}\left(t_{1} \sigma_{1} \otimes \lambda_{1}+t_{2} \sigma_{2} \otimes \lambda_{13}+t_{3} \sigma_{3} \otimes \lambda_{3}\right)\ ,\;\label{mqs24}
	\end{equation}
	where $t_{\mu} \in \mathbb{R}$ and $\sigma_{\mu},\lambda_{\nu}$ are generators of SU(2) and SU(4) respectively. If the state is separable, by definition it can then be expressed as
	$\rho_{A B}=\sum_{i=1}^{L} p_{i} \rho_{i}^{(A)} \otimes \rho_{i}^{(B)}$ with $\rho_{i}^{(A)}=\frac{1}{2} \mathds{1}+\frac{1}{2} \vec{r}_{i} \cdot \vec{\sigma}$
	and $\rho_{i}^{(B)}=\frac{1}{4} \mathds{1}+\frac{1}{2} \vec{s}_{i} \cdot \vec{\lambda}$.
	
	According to the discussion in above, the correlation matrix $\mathcal{T}=M_{r p} \cdot M_{s p}^{\mathrm{T}}$ may write explicitly as
	\begin{equation}
	\mathcal{T}=\left(\begin{array}{cccc}{\alpha_{1}} & {0} & {0} & {0} \\ {0} & {\alpha_{2}} & {0} & {0} \\ {0} & {0} & {\alpha_{3}} & {0}\end{array}\right) Q \cdot Q^{T}\left(\begin{array}{ccc}{\beta_{1}} & {0} & {0} \\ {0} & {\beta_{2}} & {0} \\ {0} & {0} & {\beta_{3}} \\ {0} & {0} & {0}\end{array}\right). \; \label{correlation2x4}
	\end{equation}
	Here $t_{\mu}=\alpha_{\mu} \beta_{\mu},$ and $Q \in \mathrm{SO}(4)$ is chosen as
	\begin{equation}
	Q=\frac{1}{2}\left(\begin{array}{cccc}
	{1} & {-1} & {-1} & {1} \\
	{-1} & {-1} & {1} & {1} \\
	{-1} & {1} & {-1} & {1} \\
	{1} & {1} & {1} & {1}\end{array}\right) ,\;
	\end{equation}
	where without loss of generality we can set $\sqrt{p_{i}}=Q_{4 i}=\frac{1}{2}$. Then the factor of separable state can be expressed as
	\begin{equation}
	M_{r p}=M_{r} D_{p}^{\frac{1}{2}}=\left(\begin{array}{cccc}
	{\alpha_{1}} & -{\alpha_{1}} & -{\alpha_{1}} & {\alpha_{1}} \\
	-{\alpha_{2}}&-{\alpha_{2}} & {\alpha_{2}} & {\alpha_{2}} \\
	{-\alpha_{3}} & {\alpha_{3}} & {-\alpha_{3}} & {\alpha_{3}}
	\end{array}\right)
	\left(\begin{array}{cccc}
	{\frac{1}{2}} & {0} & {0} & {0} \\
	{0} & {\frac{1}{2}} & {0} & {0} \\
	{0} & {0} & {\frac{1}{2}} & {0} \\
	{0} & {0} & {0} & {\frac{1}{2}}
	\end{array}\right) ,\;\label{mrp24}
	\end{equation}
	
	\begin{equation}
	M_{s p}=M_{s} D_{p}^{\frac{1}{2}}=\left(\begin{array}{cccc}
	{\beta_{1}} & -\beta_{1} & -\beta_{1} & \beta_{1} \\
	-\beta_{2} & -\beta_{2} & {\beta_{2}} & {\beta_{2}} \\
	{-\beta_{3}} & {\beta_{3}} & -\beta_{3} & \beta_{3}
	\end{array}\right)\left(\begin{array}{cccc}
	{\frac{1}{2}} & {0} & {0} & {0} \\
	{0} & {\frac{1}{2}} & {0} & {0} \\
	{0} & {0} & {\frac{1}{2}} & {0} \\
	{0} & {0} & {0} & {\frac{1}{2}}\end{array}\right).\;\label{msp24}
	\end{equation}
	(\ref{mrp24}) and (\ref{msp24}) indicate that $p_{1}=p_{2}=p_{3}=p_{4}=1 / 4$ and the Bloch vectors of $\rho_{i}^{(A, B)}$ read
	\begin{equation}
	\vec{r}_{1}=\left(\begin{array}{c}{\alpha_{1}} \\ {-\alpha_{2}} \\ {-\alpha_{3}}\end{array}\right),~~ \vec{r}_{2}=\left(\begin{array}{c}{-\alpha_{1}} \\ {-\alpha_{2}} \\ {\alpha_{3}}\end{array}\right), ~~\vec{r}_{3}=\left(\begin{array}{c}{-\alpha_{1}} \\ {\alpha_{2}} \\ {-\alpha_{3}}\end{array}\right), ~~\vec{r}_{4}=\left(\begin{array}{c}{\alpha_{1}} \\ {\alpha_{2}} \\ {\alpha_{3}}\end{array}\right)
	\end{equation}
	\begin{equation}
	\vec{s}_{1}=\left(\begin{array}{c}{\beta_{1}} \\ {-\beta_{2}} \\ {-\beta_{3}}\end{array}\right),~~ \vec{s}_{2}=\left(\begin{array}{c}{-\beta_{1}} \\ {-\beta_{2}} \\ {\beta_{3}}\end{array}\right), ~~\vec{s}_{3}=\left(\begin{array}{c}{-\beta_{1}} \\ {\beta_{2}} \\ {-\beta_{3}}\end{array}\right),~~ \vec{s}_{4}=\left(\begin{array}{c}{\beta_{1}} \\ {\beta_{2}} \\ {\beta_{3}}\end{array}\right)\ .
	\end{equation}
	The state $\rho^{(B)}=\frac{1}{4} \mathds{1}+\frac{1}{2}\left(\beta_{1} \lambda_{1}+\beta_{2} \lambda_{13}+\beta_{3} \lambda_{3}\right)$
	requires that $\beta_{1}^{2}+\beta_{3}^{2} \leq \frac{1}{4}$ and $\beta_{2}^{2} \leq \frac{1}{4}$.
	
	Therefore we may conclude that the state $\rho_{A B}$ is separable while
	$\frac{t_{1}^{2}}{\alpha_{1}^{2}}+\frac{t_{3}^{2}}{\alpha_{3}^{2}} \leq \frac{1}{4}, \frac{t_{2}^{2}}{\alpha_{2}^{2}} \leq \frac{1}{4}$ with $\alpha_{1}^{2}+\alpha_{2}^{2}+\alpha_{3}^{2} \leq 1$.
	
	{\bf Example 2:}  Application to $2\times 3\times 3$ state.
	
	Consider a $2\times 3\times 3$ tripartite quantum system, after the loss of qubit we are left with a bipartite mixed state \cite{3122}
	\begin{equation}
	\rho=\frac{1}{4}\left(I_{9}-\sum_{i=0}^{4} | \psi_{i}\rangle\langle\psi_{i} |\right)\;.
	\end{equation}
	Here, $| \psi_{0} \rangle=| 0 \rangle( | 0\rangle-| 1 \rangle ) / \sqrt{2}$, $| \psi_{1} \rangle=( | 0\rangle-| 1 \rangle ) | 2 \rangle / \sqrt{2}$,
	$| \psi_{2} \rangle=| 2 \rangle( | 1\rangle-| 2 \rangle ) / \sqrt{2}$, $| \psi_{3} \rangle=( | 1\rangle-| 2 \rangle ) | 0 \rangle / \sqrt{2}$,
	$| \psi_{4} \rangle=( | 0\rangle+| 1 \rangle+| 2 \rangle )( | 0\rangle+| 1 \rangle+| 2 \rangle ) / 3$.
	So in the Bloch representation, we find
	
	\begin{equation}
	T=-\frac{1}{4}\left(
	\begin{array}{rrrrrrrr}
	1 & 0 & 0 & 1 & 0 & 1 & 0 & \frac{\sqrt{27}}{2} \\
	0 & 0 & 0 & 0 & 0 & 0 & 0 & 0 \\
	-\frac{9}{4} & 0 & -\frac{9}{8} & 0 & 0 & 0 & 0 & \frac{\sqrt{27}}{8} \\
	1 & 0 & 0 & 1 & 0 & 1 & 0 & 0 \\
	0 & 0 & 0 & 0 & 0 & 0 & 0 & 0 \\
	1 & 0 & -\frac{9}{4} & 1 & 0 & 1 & 0 & -\frac{\sqrt{27}}{4} \\
	0 & 0 & 0 & 0 & 0 & 0 & 0 & 0 \\
	-\frac{\sqrt{27}}{4} & 0 & \frac{\sqrt{27}}{8} & 0 & 0 & \frac{\sqrt{27}}{2} & 0 & -\frac{3}{8} \\
	\end{array}
	\right),
	\end{equation}
	Here $||T||_{KF}\simeq3.1603$, and hence, as per condition (\ref{separable-kfnorm}) the $3\times 3$ is an entangled bipartite
	mixed state \cite{lq18b,vi07}, which means $2\times 3\times 3$ is a robust tripartite state.
	
	\begin{observation}
		A tripartite quantum state $|\Psi\rangle_{2\times N\times M}$ shall be rebust if the reduced density matrix $\rho^{NM}$ is a convex combination of $|\psi^{\pm}\rangle=\alpha|ee^\bot\rangle\pm\beta|e^\bot e\rangle$ and normalized identity matrix with probability $P(|\psi^{\pm}\rangle)> \frac{1}{N+1}$ and nonzero coefficients $\alpha\neq\beta$.
		\label{observ-3}
	\end{observation}

	{\bf Example 3:} Consider the following $2\times N\times M$ quantum state:
	\begin{equation}
	|\phi\rangle=\frac{1}{\sqrt\alpha}(\beta_1|0NM\rangle+\beta_2|10M\rangle+\beta_3|1N0\rangle) \; .
	\end{equation}
	Here $\alpha $ is a normalization constant. If the qubit decouples from the other two particles, the remaining state in
	density matrix form reads
	\begin{equation}
	\rho^{NM}=p|\psi\rangle\langle\psi|+(1-p)|NM\rangle\langle NM| \; ,
	\end{equation}
	where $|\psi\rangle = \frac{1}{\sqrt 2}(|0M\rangle+|N0\rangle)$ is a maximally entangled bipartite state. As long as $p\neq 0$,
	$\rho^{MN}$ will remain to be an entangled state.
	
	\begin{observation}
		Provide the bipartite reduced density matrix partial transpose contains at least one negative eigenvalue, all tripartite quantum states $|\Psi\rangle_{2\times N\times M}$ will be robust ones with the loss of qubit.
		\label{ppt-3}
	\end{observation}
	
	{\bf Example 4:} Consider a pure $2\times 3\times 3$ quantum state
	\begin{equation}\label{3ps}
	|\psi\rangle=\frac{1}{\sqrt{4}}(|010\rangle+|001\rangle+|112\rangle+|121\rangle),
	\end{equation}
	the loss of qubit returns the reduced density matrix
	\begin{equation}\label{rho-9p}
	\rho^{AB}= |10\rangle\langle10|+|10\rangle\langle01|+|01\rangle\langle10|+|01\rangle\langle01|+
	|12\rangle\langle12|+|12\rangle\langle21|+|21\rangle\langle12|+|21\rangle\langle21|.
	\end{equation}
	The partial transpose of $\rho^{AB}$ with respect to the first particle takes the following form
	\begin{equation}
	\rho^{T_\mathcal{A}}=\frac{1}{4}
	\left(
	\begin{array}{ccccccccc}
	0 & 0 & 0 & 0 & 1 & 0 & 0 & 0 & 0 \\
	0 & 1 & 0 & 0 & 0 & 0 & 0 & 0 & 0 \\
	0 & 0 & 0 & 0 & 0 & 0 & 0 & 0 & 0 \\
	0 & 0 & 0 & 1 & 0 & 0 & 0 & 0 & 0 \\
	1 & 0 & 0 & 0 & 0 & 0 & 0 & 0 & 1 \\
	0 & 0 & 0 & 0 & 0 & 1 & 0 & 0 & 0 \\
	0 & 0 & 0 & 0 & 0 & 0 & 0 & 0 & 0 \\
	0 & 0 & 0 & 0 & 0 & 0 & 0 & 1 & 0 \\
	0 & 0 & 0 & 0 & 1 & 0 & 0 & 0 & 0 \\
	\end{array}
	\right).
	\end{equation}
	The $\rho^{T_\mathcal{A}}$ has eigenvalues $\left\{-\frac{1}{2 \sqrt{2}},\frac{1}{2 \sqrt{2}},\frac{1}{4},\frac{1}{4},\frac{1}{4},\frac{1}{4},0,0,0\right\}$, in which
	the first eigenvalue is negative. Then according to the PPT criterion \cite{hlvc00}, the bipartite
	mixed state $\rho^{AB}$ is still an entangled state. Therefore, the pure $2\times 3\times 3$ tripartite state
	is a robust quantum state with the loss of qubit.
	
	\section{Measuring the robustness of quantum State}
	
	Above, we have identified robust and fragile tripartite states $|\Psi\rangle_{2\times N\times M}$
	with respect to the loss of qubit. Their robustness can be estimated by using an
	entanglement measure for the residual mixed state \cite{nmb18}.
	The entanglement entropy \cite{petz01} and R\'{e}nyi entropy \cite{nobili17} can measure entanglement of
	pure bipartite systems,  but for mixed states the entropy is not a good means for such an aim, as it is sensitive
	to both quantum and classical correlations.
	To investigate the robustness of the states considered  here, we evaluate explicitly the dynamics of the negativity \cite{vw02,lee03} and the concurrence \cite{w98}.

	For the residual mixed states $\rho\in\mathcal{H}_{A}\otimes\mathcal{H}_{B}$, the negativity is defined as the absolute value of the sum of the negative eigenvalues of the partially transposed (PT) density matrix $\rho^{T_{\mathcal{A}}}_{AB}$, that can be defined in terms of the trace norm $\|\rho^{T_{\mathcal{A}}}_{AB}\|$ as
	\begin{equation}
	\mathcal{N}(\rho)=\frac{ \|\rho^{T_{\mathcal{A}}}_{AB}\|-1}{2}\ .\;\label{negativity}
	\end{equation}
	Here $\rho^{T_{\mathcal{A}}}_{AB}$ is partial transpose of $\rho$ with respect to
	partite $\mathcal{A}$ and $\|\rho\|=Tr(\sqrt{\rho\rho^\dagger})$. Note, generally speaking,
	the negativity relation fails in measuring the entanglement of those entangled states with dimensions higher than six \cite{rl10}. Though the positivity of partial transpose is a necessary and sufficient condition for non-distillability in $2\otimes n$ quantum systems \cite{dur00,horodecki97}, there still exists some entangled states with positive partial transpose (PPT) in bipartite systems
	with dimensions larger than six, which are usually named the PPT bound entangled states. Despite the negativity expression (\ref{negativity}) cannot measure the robustness of PPT bound entangled states, the convex-roof extended negativity (CREN) \cite{lee03} may compute robustness of a high dimensional residual bipartite state. The negativity vanishes only when the state is fragile, i.e. the null negativity means residual state cannot carry information anymore.
	
\begin{figure}\centering
		\scalebox{0.5}{\includegraphics{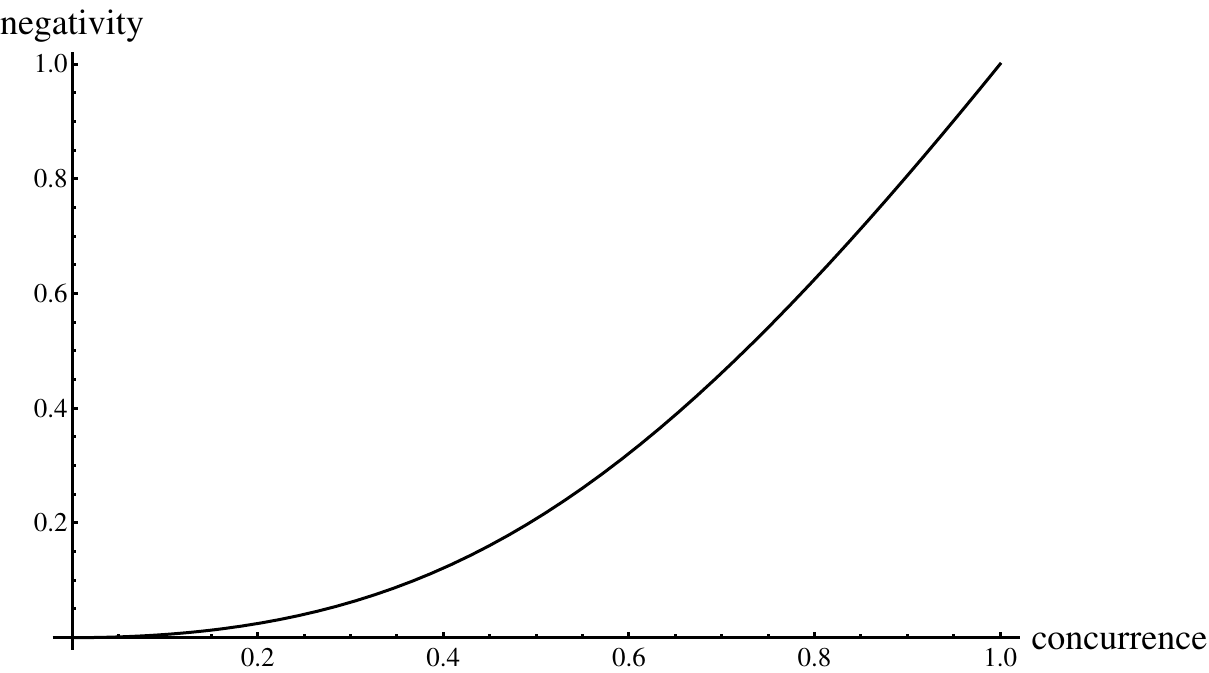}}
		\caption{Illustration of the negativity does not exceed the concurrence.} \label{fig-nc}
	\end{figure}

	The concurrence \cite{w98} for a bipartite pure state $|\psi\rangle$ is defined as
	\begin{equation}
	C(|\psi\rangle)= \sqrt{2(1-tr\rho^2_{A})}\ ,
	\end{equation}
	where $\rho_A=tr_B(\rho_{AB})$. This measure can be extended to the mixed states
	$\rho$ by virtue of a convex roof construction \cite{fma04}, that is
	\begin{equation}
	C(\rho)=\underset{\{p_i,|\psi_i\rangle\}}{\text{inf}}\sum_ip_iC(|\psi_i\rangle)
	\end{equation}
	with $p_i \geq 0$ the probability of various pure states $|\psi_i\rangle$ in the mixed state $\rho = \sum_ip_i|\psi_i\rangle\langle\psi_i|$. The $C(\rho)$ may vanish if and only if $\rho$ is fragile. Fig.\ref{fig-nc} shows the relation between concurrence and negativity.
	
	\section{Conclusion}
	
	In quantum information processing with multiple entangled particles, to keep the processing flow in the presence of certain environmental interference, e.g. some parties decouple from the system, is an extremely important question. Robust quantum states have the ability to keep on transferring information even with the loss of some parties. In this work, we investigated the robustness of $2\times N\times M$ quantum states with respect to the loss of the qubit in terms of negativity and concurrence of the bipartite reduced density matrix. We found that after the removal of the qubit from $2\times N\times M$ tripartite quantum states, if the bipartite reduced density matrices are not reducible according to local rank, the tripartite quantum states are robust, otherwise are fragile. We identified the fragility of this type of quantum states in terms of a correlation matrix $\mathcal{T}$ decomposition under some constraints. In the future, to understand the nature of a chain of entangled particles in arbitrary dimension is highly expected.
	
	\vspace{.7cm} {\bf Acknowledgments} \vspace{.3cm}
	
	This work was supported in part by the National Natural Science Foundation of China(NSFC) under the Grants 11975236 and 11635009. S.M.Z. was supported in part by the CAS-TWAS Fellowship. We are also grateful to J. L. Li for some fruitful discussions on this work.
	


\begin{thebibliography}{99}
\bibitem{3122} This example is taken from ref. \cite{vi07}, we recommend  \cite{vi07} and \cite{bdmsst99} to clarify basic concepts.
\end{thebibliography}
\end{document}